\documentclass[PAPER, atlasdraft=false, cernpreprint, UKenglish, texmf, orcidlogo]{atlasdoc}
\usepackage[subcaption,subfig,default]{atlaspackage}
\usepackage{atlasbiblatex}

\usepackage{atlasphysics}
\usepackage{atlasheavyion}
\usepackage{atlasjetetmiss}

\graphicspath{{logos/}{figures/}}

\usepackage{ANA-HION-2020-06-PAPER-defs}

\AtlasTitle{Measurement of substructure-dependent suppression of large-radius jets with charged particles in Pb+Pb collisions with ATLAS}

\AtlasAbstract{%
Measurements of jet substructure in Pb+Pb collisions provide key insights into the mechanism of jet quenching in the hot and dense QCD medium created in these collisions. This Letter presents a measurement of the suppression of large-radius jets with a radius parameter of $R = 1.0$ and its dependence on the jet substructure. The measurement uses 1.72~nb$^{-1}$ of Pb+Pb data and 255~pb$^{-1}$ of $pp$ data, both at $\sqrt{s_{_\mathrm{NN}}} = 5.02$~TeV, recorded with the ATLAS detector at the Large Hadron Collider. Large-radius jets are reconstructed by reclustering $R = 0.2$ calorimetric jets and are measured for transverse momentum above $200$~GeV. Jet substructure is evaluated using charged-particle tracks, and the overall level of jet suppression is quantified using the jet nuclear modification factor ($R_\mathrm{AA}$). The jet $R_\mathrm{AA}$ is measured as a function of jet $p_{\mathrm{T}}$, the charged $k_t$ splitting scale ($\sqrt{d_{12}}$), and the angular separation ($\Delta R_{12}$) of two leading sub-jets.
The jet $R_\mathrm{AA}$ gradually decreases with increasing $\sqrt{d_{12}}$, implying significantly stronger suppression of large-radius jets with larger $k_t$ splitting scale.
The jet $R_\mathrm{AA}$ gradually decreases for $\Delta R_{12}$ in the range $0.01{-}0.2$ and then remains consistent with a constant for $\Delta R_{12} \gtrsim 0.2$.
The observed significant dependence of jet suppression on the jet substructure will provide new insights into its role in the quenching process.
}

\AtlasRefCode{HION-2020-06}

\PreprintIdNumber{CERN-EP-2025-078}
\AtlasJournalRef{Phys. Lett. B. 871 (2025) 139929}
\AtlasDOI{10.1016/j.physletb.2025.139929}

\AtlasJournal{Phys.\ Lett.\ B.}

\hypersetup{pdftitle={ATLAS document},pdfauthor={The ATLAS Collaboration}}

\begin{document}

\maketitle

\section{Introduction}

Collisions of heavy ions at the Relativistic Heavy Ion Collider (RHIC) and the Large Hadron Collider (LHC)~\cite{Evans:2008zzb} allow one to study the interaction between showers of high transverse momentum ($\pt$) partons produced in hard-scattering processes and quark--gluon plasma (QGP) formed as a result of copious liberation of colour charges during the nuclear collisions~\cite{Busza:2018rrf,Shuryak:2014zxa}.
The parton showers, which can be described within perturbative quantum chromodynamics, give rise to jets of hadrons that can be measured in a detector.
The interactions between the parton shower and the QGP can reduce the energies of jets and modify
their substructure~\cite{Casalderrey-Solana:2007knd,dEnterria:2009xfs,Wiedemann:2009sh,Majumder:2010qh,Mehtar-Tani:2013pia,Cao:2020wlm}.
These jet modifications
are
referred to as \textit{jet quenching}
and are measured with a variety of observables at RHIC and the LHC~\cite{Connors:2017ptx,Cunqueiro:2021wls,Apolinario:2024equ}.
Developing a quantitative theoretical description of jet quenching is an open problem that requires detailed experimental input.

One of the important mechanisms contributing to the modification of jet substructure is coherent radiation.
When the parton shower propagates through the QGP medium,
colour coherence makes the medium-induced gluon radiation unable to resolve small-angle splittings of the parton shower. This implies that a part of the parton shower containing small-angle splittings radiates coherently as a single colour charge~\cite{Mehtar-Tani:2010ebp,Mehtar-Tani:2011vlz,Mehtar-Tani:2011hma}. Consequently, the colour coherence leads to an overall smaller energy loss for smaller-angle splittings than for the case of larger-angle splittings that interact incoherently with the QGP~\cite{Casalderrey-Solana:2012evi}. To describe this phenomenon, a parameter quantifying what part of the collimated shower is resolved needs to be introduced. Typically, this parameter is the resolution length \lres~\cite{Hulcher:2017cpt} or, in an approximation of multiple soft scatterings in QGP, the (de)coherence angle $\theta_\text{c}$~\cite{Mehtar-Tani:2010ebp,Andrews:2018jcm}. Two partons in the shower separated by a distance or angle smaller than \lres or $\theta_\text{c}$ are not resolved by the medium.

Several recent measurements using various jet substructure observables~\cite{Andrews:2018jcm,Apolinario:2024equ} have provided experimental input that may help to quantify the resolution parameter. In particular, these include measurements of splitting functions, groomed jet radius, and jet girth~\cite{CMS-HIN-16-006,ALICE:2021mqf,CMS-HIN-23-001,HION-2021-09,ALICE:2019ykw,HION-2019-09}; measurements of the number of splittings and the $N$-subjettiness~\cite{ALICE:2019ykw,ALICE:2021vrw}; and double-differential measurements of jet fragmentation~\cite{HION-2018-03}.
To isolate sub-jets entering the calculation of substructure observables, two strategies are used in these measurements. The first strategy relies on reclustering the constituents of jets~\cite{Butterworth:2008iy} and employing a grooming method, such as Soft Drop~\cite{Marzani:2019hun,Larkoski:2014wba}.
The second strategy relies on clustering jets with small radii ($R=0.2$) that satisfy a minimum-$\pt$ requirement into a jet with large radius ($R=1.0$). The latter strategy was used in Ref.~\cite{HION-2019-09},
where a significantly weaker suppression of reclustered $R=1.0$ jets composed of single sub-jets was seen in comparison with reclustered $R=1.0$ jets composed of multiple sub-jets.
Reclustered $R=1.0$ jets are used also in the study presented here and they are called \lR jets in the subsequent text.

A recent theoretical study~\cite{Kudinoor:2025ilx}, which analysed the
response of the QGP to the passing jet within the Hybrid strong/weak
coupling model~\cite{Casalderrey-Solana:2014bpa}, demonstrated that \lR-jet substructure measurements may be used to directly constrain \lres.
However, that study also shows that the coarse binning of substructure observables used in Ref.~\cite{HION-2019-09}, due to the use of calorimetric objects only, does not allow precise quantification of \lres. This limitation is a consequence of the fact that two $R=0.2$ jets are separated by a minimum angular distance of approximately 0.2. To overcome this limitation, objects possessing better angular resolution, namely charged-particle tracks, may be used to study jet substructure. The approach with charged-particle tracks matched to calorimeter energy clusters was previously used in the substructure study in Ref.~\cite{HION-2021-09}, where grooming was applied to jets with the traditional radius of $R=0.4$.
The measurement presented here uses large-$R$ jets together with charged-particle tracks to study jet substructure over the broad range of angular separation between sub-jets. The new measurement thus connects the high angular resolution of charged-particle tracks with the large phase-space coverage and precision of \lR calorimetric jets. This approach should, therefore, help to quantify the impact of colour coherence on the jet energy loss.

The measurement presented in this Letter uses data collected by the ATLAS experiment in \PbPb\ and \pp\ collisions at a centre-of-mass energy of 5.02~TeV per colliding nucleon pair. The \PbPb\ and \pp\ data samples contain 1.72~nb$^{-1}$ and 255~pb$^{-1}$ of collision data, respectively.


%

%
%
\section{Experimental set-up}

\newcommand{\AtlasCoordFootnote}{%
ATLAS uses a right-handed coordinate system with its origin at the nominal interaction point (IP) in the centre of the detector and the \(z\)-axis along the beam pipe.
The \(x\)-axis points from the IP to the centre of the LHC ring, and the \(y\)-axis points upwards. Polar coordinates \((r,\phi)\) are used in the transverse plane,
\(\phi\) being the azimuthal angle around the \(z\)-axis. The pseudorapidity is defined in terms of the polar angle \(\theta\) as \(\eta = -\ln \tan(\theta/2)\) and is equal to the rapidity $ y = (1/2) \ln[(E + p_z)/(E - p_z)] $ in the relativistic limit. Angular distance is measured in units of \(\Delta R \equiv \sqrt{(\Delta y)^{2} + (\Delta\phi)^{2}}\).}

The ATLAS experiment~\cite{PERF-2007-01} at the LHC is a multipurpose particle detector with a forward--backward symmetric cylindrical geometry and  nearly \(4\pi\) coverage in solid angle.\footnote{\AtlasCoordFootnote}
It consists of an inner tracking detector surrounded by a thin superconducting solenoid providing a \qty{2}{\tesla} axial magnetic field, electromagnetic and hadronic calorimeters, and a muon spectrometer.

The inner tracking detector covers the pseudorapidity range \(|\eta| < 2.5\). It consists of silicon pixel, silicon microstrip, and transition radiation tracking detectors. The calorimeter system consists of a sampling liquid-argon (LAr) electromagnetic (EM) calorimeter covering $|\eta| < 3.2$, a steel--scintillator sampling hadronic calorimeter covering $|\eta| < 1.7$, LAr hadronic calorimeters covering $1.5 < |\eta| < 3.2$, and two LAr forward calorimeters (FCal) covering $3.1 < |\eta| < 4.9$.
The endcap and forward regions are thus instrumented with LAr calorimeters
for EM and hadronic energy measurements up to \(|\eta| = 4.9\).
A zero-degree calorimeter (ZDC) was situated at $|\eta| > 8.3$ during Pb+Pb data-taking. It is composed of alternating layers of quartz rods and tungsten plates, and is mostly sensitive to spectator neutrons from fragmenting nuclei in Pb+Pb collisions.

A two-level trigger system is used to select events~\cite{TRIG-2016-01}.
The first-level trigger is implemented in hardware and uses a subset of the detector information
to accept events at a rate close to 100~kHz. This is followed by a software-based trigger that reduces the accepted rate of complete events to 1.25~kHz on average
depending on the data-taking conditions.
A software suite~\cite{SOFT-2022-02} is used in data simulation, in the reconstruction
and analysis of real and simulated data, in detector operations, and in the trigger and data acquisition systems of the experiment.


%

%
\section{Data and Monte Carlo selection}

The \PbPb\ and \pp\ data used in these measurements were collected in 2018 and 2017~\cite{DAPR-2018-01}, respectively.
Events were selected using a combination of calorimeter-based jet triggers~\cite{ATL-DAQ-PUB-2019-001,TRIG-2016-01}. Both the \pp\ and \PbPb\ events are required to contain at least one primary vertex reconstructed within 150~mm of the nominal interaction point along the beam axis~\cite{ATL-PHYS-PUB-2015-026}. Only events taken during stable beam conditions and satisfying detector and data-quality requirements, which include normal operation of the calorimeters and inner tracking detectors, are considered. All jets in the analysis are in a kinematic range where the jet trigger is fully efficient. The Pb+Pb data contain only a small fraction of events $(<0.5\%)$ with multiple collisions per bunch crossing, and these are suppressed using the anti-correlation of signals from the ZDC and FCal. The \pp\ data were collected with typically $1.4{-}4.4$ inelastic interactions per bunch crossing. No \pileup rejection is applied in the analysis of \pp\ data.

The level of overall event activity, or \emph{centrality}, in \PbPb\ collisions is characterized using the total transverse energy in the forward calorimeter, $\Sigma E_\mathrm{T}^\mathrm{FCal}$.
The $\Sigma E_\mathrm{T}^\mathrm{FCal}$ distribution is divided into percentiles of the total inelastic cross-section for \PbPb\ collisions.
Events in different ranges of $\Sigma E_\mathrm{T}^\mathrm{FCal}$ are associated with an underlying \PbPb\ collision geometry according to a Monte Carlo (MC) Glauber simulation which also allows us to define the nuclear thickness function, $\langle\TAA\rangle$~\cite{Miller:2007ri,HION-2024-03}.
This measurement uses five centrality intervals
corresponding to the following
$\Sigma E_\mathrm{T}^\mathrm{FCal}$ percentiles, ordered from the most central to the most peripheral collisions (highest to lowest $\Sigma E_\mathrm{T}^\mathrm{FCal}$): $0{-}10\%$, $10{-}20\%$, $20{-}40\%$, $40{-}60\%$, and $60{-}80\%$.

The performance of the detector and of the analysis procedure for \pp\ collisions is evaluated using $3.2 \times 10^7$ MC events. This simulation sample comprises \PYTHIA[8.2]~\cite{Sjostrand:2014zea} dijet events generated at a centre-of-mass energy of $\sqrt{s} = 5.02$~TeV with the A14 set of tuned parameters~\cite{ATL-PHYS-PUB-2014-021} and the \NNPDF[2.3lo] parton distribution functions (PDFs)~\cite{Ball:2012cx}.
\Pileup from additional \pp\ collisions was generated by \PYTHIA[8.1]~\cite{Sjostrand:2007gs}, with the same PDFs and the parameter values set to the A3 tune~\cite{ATL-PHYS-PUB-2016-017}, and the distribution of the number of extra collisions was matched to that of data.
A separate sample of $3.2 \times 10^7$ simulated events was produced to evaluate the performance of the detector and analysis procedure for \PbPb\ collisions. This simulation sample was produced from minimum-bias Pb+Pb data events overlaid with hard-scattering \PYTHIA[8] dijet \pp\ events generated at a centre-of-mass energy of $\sqrt{s} = 5.02$~TeV with the same tune and PDFs as used in generating the \pp\ MC sample for \pp\ collisions. This procedure accurately reproduces the impact of the underlying event on track reconstruction and the detector's response to jets. This sample is reweighted on an event-by-event basis to ensure that
the $\Sigma E_\mathrm{T}^\mathrm{FCal}$ distribution is the same
as is measured in the data sample. The detector's response in both samples was simulated utilizing \GEANT{}~\cite{GEANT4:2002zbu,SOFT-2010-01}.


%

%
\section{Jet and track reconstruction}
\label{sec:jettrack}

Two types of objects from the event reconstruction enter the analysis:
calorimetric jets and charged-particle tracks from the inner detector.

Jets are reconstructed using the \antikt\ algorithm~\cite{Cacciari:2008gp,Fastjet} with a radius parameter of $R=0.2$. The jet reconstruction procedure follows that used previously in \PbPb\ collisions~\cite{HION-2017-10,HION-2011-02}, and is summarized here. Calorimeter cells in all calorimeter layers, with their energy at the electromagnetic scale, are grouped into $\Delta\eta \times \Delta\phi = 0.1 \times \pi/32$ logical towers. The towers are used as the input to the initial jet-finding performed using the $R=0.2$ \antikt\ algorithm. After the initial jet-finding, the contribution from the underlying event (UE) to the energy deposited in the towers is estimated on an event-by-event basis.
This is done by averaging over energies deposited in towers at a given $\eta$ position. During the averaging and subsequent subtraction, the $\phi$ dependence arising from the global collective flow in Pb+Pb collisions is also taken into account.
Information from towers within $\Delta R = 0.4$ of jet candidates is excluded from the estimate of UE energy. The energies in towers inside jets are updated to subtract the estimated UE contribution, and the UE procedure is iterated using a better-defined set of jets to define the exclusion regions. The resulting jet kinematics are corrected using $\pt$- and $\eta$-dependent factors, determined from simulation, to account for the response of the calorimeter to jets %
\cite{JETM-2018-05}.
An additional correction for the absolute response in data is based on \textit{in situ} studies of \pp\ collisions where jets recoil against photons, $Z$ bosons, or jets in other regions of the calorimeter~\cite{JETM-2018-05}. This calibration is followed by a cross-calibration that relates the jet energy scale (JES) in high-luminosity 13~TeV \pp\ collisions~\cite{ATLAS-CONF-2015-016}, used for \textit{in situ} studies, to that of jets reconstructed in the 5.02~TeV data using the procedure outlined above.     %
The same jet reconstruction procedure, without the azimuthal modulation of the UE, is also applied to \pp\ collisions. The UE subtraction in \pp\ collisions removes the \pileup contribution to the jet.

Charged-particle tracks are reconstructed from hits in the inner detector using the standard ATLAS track reconstruction algorithm with settings optimized for the high hit density in heavy-ion collisions~\cite{PERF-2015-08}. Only tracks with $\pt$ greater than 4~\GeV are used in this analysis.
A requirement on the minimum number of hits and maximum number of holes, where a hole is defined as the absence of a hit expected to be on the track trajectory, ensures the high quality of reconstructed tracks.
In order to suppress the contribution from secondary particles, requirements are imposed on the distance of closest approach of the track to the primary vertex in the transverse ($d_0$) and longitudinal (\zzsth) directions.
These requirements vary with \pT\ and $\eta$, but they do not exceed 0.3 mm for $d_0$ and 0.5 mm for \zzsth at \pT\ = 4~\GeV.
The track reconstruction efficiency is approximately 80\% and 85\% in \PbPb\ and \pp\ collisions, respectively. The multiplicity of charged-particle tracks with $\pt>4$~\GeV in \lR jets with $200<\pt<251$~\GeV reaches a value of approximately 7 in \pp\ collisions.


%
%
%
\section{Analysis procedure}

The reconstruction of large-$R$ jets follows the procedure introduced in Ref.~\cite{HION-2019-09}. First, $R=0.2$ \antikt\ jets are reconstructed in the calorimeter using the method described in Section~\ref{sec:jettrack}. The $R=0.2$ jets with $\pt > 35$~\GeV\ and $|\eta|<3.0$ are then reclustered using the \antikt\ algorithm with radius parameter $R=1.0$ to form so-called \lR jets. This approach enables measurements over a broad kinematic range, mitigates the impact of the UE, prevents possible biases from the recovery of quenched-jet energy that is transferred outside the $R=0.2$ sub-jets, and reduces systematic uncertainties~\cite{HION-2019-09}. This approach also allows to define sub-jets using the information from the reclustering of $R=0.2$ jets and to calculate substructure observables of interest. The use of $R=0.2$ sub-jets, however, suffers from the limitation on the minimum angular separation between the sub-jets, which is close to $0.2$. To overcome this limitation, charged particles reconstructed as tracks in the inner detector are used in this measurement to define sub-jets of \lR jets, without changing the kinematics of \lR jets. Details of the sub-jet definition are given in the next paragraph.

First, charged-particle tracks with $\pt>4$~\GeV\ that satisfy the requirements specified in Section~\ref{sec:jettrack} are matched to $R=1.0$ jets. The track-to-jet matching is done by \textit{ghost-association}~\cite{Cacciari:2007fd,Cacciari:2008gn}, where tracks are treated as being infinitesimally soft, with $\pt=1$~eV. The tracks with this replaced \pt\ are added to the list of inputs together with the $R=0.2$ jets to be clustered into $R=1.0$ jets with the \antikt\ algorithm. This allows the matching to proceed, while the infinitesimally low \pt\ of tracks implies that they do not affect the reconstruction of large-$R$ jets.
Tracks with their original kinematics associated with a $R=1.0$ jet are then reclustered using the \kt\ clustering algorithm~\cite{Catani:1993hr} with $R=2.5$. While the majority of tracks would be clustered with a radius parameter $R=1.0$, the larger radius parameter is used in the \kt\ algorithm to cover edge effects and ensure that all associated tracks are clustered into one jet.
The reclustering is performed with the Soft Drop grooming condition applied with parameters $z_\mathrm{cut} = 0.15$, which optimizes the suppression of the UE contribution, and $\beta=0$, which guarantees that the grooming has no impact on the angular structure of the jet; applying Soft Drop condition does not invoke a separate running of Cambridge/Aachen algorithm. Approximately 5\% of the jets fail to satisfy the Soft Drop condition and are not considered in the analysis.
Furthermore, the ratio of the $\Sigma\pt$ of tracks associated with a jet to the reconstructed \pt\ of the calorimetric jet is required to be greater than 0.2 because the absence of charged particles would undermine meaningful substructure evaluation. This requirement also enhances the robustness of the unfolding procedure which is defined later. The same requirement is applied to particle-level jets. The coverage of the inner tracking detector, along with the choice of radius parameter $R$ for track reclustering, limits the pseudorapidity selection requirement for large-$R$ jets to $|\eta|<1.3$.

The reclustering gives access to the two leading track-based sub-jets of each \lR jet and defines their angular separation, $\dRonetwo$, and $\kt$ splitting scale, $\Donetwo$, as
\begin{equation}
\nonumber
\dRonetwo  = \sqrt{ (\Delta y_{12})^2 + (\Delta \phi_{12})^2}, \quad
\Donetwo = \mathrm{min}({\pT}_{1},{ \pT}_{2}) \times \dRonetwo.
\end{equation}
The $\dRonetwo$ and $\Donetwo$ were used in the previous study \cite{HION-2019-09} and are directly sensitive to the medium resolution and hardness or virtuality of the splitting. Due to the Soft Drop condition with $z_\mathrm{cut}$, the phase space of \dRonetwo and \Donetwo is correlated, as very soft splittings at large angular separation are removed by grooming.

The $R=0.2$ jets are defined at the particle level in the MC sample before detector simulation by applying the anti-\kt algorithm with $R=0.2$ to stable particles with a proper lifetime greater than 30~ps
but excluding muons and neutrinos, which do not leave significant energy deposits in the calorimeter. The same kinematic requirements, the same reclustering procedure, and the same ghost-association procedure to match particles to jets that are applied to detector-level jets are also applied at the particle level to obtain particle-level jets in simulation.

To quantify the magnitude of quenching effects, the jet nuclear modification factor \RAA of large-$R$ jets is evaluated. It is defined as the ratio of per-event jet yields measured in \PbPb\ collisions normalized by $\langle\TAA\rangle$ to the jet cross-section measured in \pp\ collisions~\cite{HION-2013-06}.
The large-$R$-jet yields in \PbPb\ collisions, cross-sections in \pp\ collisions, and jet \RAA\ are measured differentially in jet \pt, and $\dRonetwo$ or $\Donetwo$.
The measurement is performed for jets with transverse momentum greater than $200$~\GeV.
This ensures that UE fluctuations do not affect the sub-jet reconstruction and kinematics.
The binning chosen for the measured distributions provides an optimal balance between statistical precision and resolution effects,
and also follows that used in an earlier measurement of \lR jets~\cite{HION-2019-09} to allow a direct comparison.

The jet reconstruction performance for \lR jets is evaluated using MC simulations. The average ratio of the reconstructed jet \pt to the corresponding particle-level jet \pt is found to be unity within 3\%. The inclusive and differential distributions are corrected for detector effects using the iterative Bayesian unfolding method~\cite{DAgostini:1994fjx,Adye:2011gm} in one and two dimensions, respectively, to reproduce the particle-level distributions. This correction accounts for bin migrations due to the jet energy resolution (JER) and JES, as well as combinatorial sub-jet contributions arising from UE fluctuations and additional hard partonic interactions in \PbPb\ collisions. To better model the data, simulated distributions are reweighted as a function of the reconstruction-level jet $\pt$, $\Donetwo$, and $\dRonetwo$ using the reconstruction-level data-to-simulation ratio prior to unfolding.
The number of unfolding iterations is chosen to ensure stability of the results while minimizing statistical uncertainty amplification. Four iterations are used for inclusive $R=0.2$ and $R=1.0$ jet yields and cross-sections, while six and eight iterations are applied to $\Donetwo$ and $\dRonetwo$ distributions in the two-dimensional unfolding, respectively. Particle-level jets that do not match a reconstructed jet satisfying the selection criteria are incorporated as an inefficiency correction after unfolding.
A correction is also applied to account for the migration in rapidity, i.e.\ for the presence of jets reconstructed within a selected rapidity region that have associated particle-level jets with rapidity outside of that region, and vice versa.


%

%
\section{Systematic uncertainties}
\label{sec:uncert}

The following sources of systematic uncertainty are considered: JES, JER, the sensitivity of the unfolding to the prior, and tracking-related and overall normalization uncertainties. For each variation accounting for a source of systematic uncertainty, all unfolded distributions, along with their \RAA\ ratios, are re-evaluated. The difference between the varied and nominal distributions is used as an estimate of the uncertainty.

The systematic uncertainties from the JES and JER are evaluated by varying the reconstructed jet energy in simulation. The JES uncertainty is split into three parts: 1) a centrality-independent baseline determined from calorimetric studies~\cite{JETM-2018-05} plus relative energy-scale differences between heavy-ion and 13~TeV \pp\ jet reconstruction~\cite{ATLAS-CONF-2015-016}; 2) additional components accounting for the quark/gluon composition of parton showers~\cite{HION-2017-10}; and 3) a centrality-dependent component accounting for a different calorimetric response to jets with a modified parton shower due to quenching in \PbPb\ collisions~\cite{ATLAS-CONF-2015-016, HION-2017-10}. The last component is evaluated by comparing the calorimetric \pt\ with the summed \pt\ of charged particles within the jet, and reaches about  1\% in central events.
The JER  uncertainty is assessed by repeating the unfolding with additional Gaussian smearing of the particle-level jet \pt, where the smearing factor is derived from dijet energy-balance studies in 13~TeV \pp\ data~\cite{JETM-2018-05} and includes additional contributions to account for the different heavy-ion jet reconstruction procedure.

Uncertainties related to track reconstruction and selection arise from several sources. First, the uncertainty due to the detector material description is evaluated by varying the material model in simulation; this can affect the track reconstruction efficiency by approximately 0.5--1.8\% in the \pt\ range used in this analysis~\cite{ATL-PHYS-PUB-2015-051}. Second, the uncertainty associated with track selection is determined by comparing the results of using the default criteria with those of using different track-quality selections. This leads to variations in the fractions of primary, fake, and secondary tracks and to different tracking efficiencies. Third, the uncertainty due to fake and secondary tracks is estimated as 100\% of their contribution determined in MC simulation. Finally, an additional 0.4\% uncertainty is assigned to account for the reduced tracking efficiency in regions of high local track density within jet cores~\cite{ATL-PHYS-PUB-2016-007}.

The systematic uncertainty in the unfolding procedure is assessed by constructing response matrices without applying the reweighting used to match the shapes of the charged-particle and jet distributions observed in the data.

Finally, there are uncertainties in the overall normalization of the measured distributions. This includes the uncertainty of 1.0\% in the determination of the \pp\ luminosity~\cite{DAPR-2021-01,LUCID2} and uncertainties of
0.5\%, 0.9\%, 1.8\%, 3.9\%, and 7.1\%
in the determination of $\langle\TAA\rangle$ values~\cite{HION-2017-10}
for successive centrality intervals from the most central to the most peripheral.
The leading systematic uncertainties affecting the yields arise from the JES, at approximately 10\%, followed by an approximately 5\% contribution from JER uncertainties. Uncertainties from track reconstruction and unfolding are smaller, typically below 2\% or 3\%, respectively. The magnitude of these uncertainties shows a mild dependence on both \Donetwo and \Ronetwo, typically increasing slightly as these variables approach either their highest or lowest values. When evaluating the uncertainty in \RAA, the uncertainties common to both \pp and \PbPb collisions, namely the centrality-independent JES, JER, and tracking uncertainties, are treated as being correlated.
Thus, the uncertainty in \RAA\ is typically almost halved relative to uncertainties in the yields.


%

%
\section{Results}

\begin{figure}[t]
\begin{subfigure}[b]{0.49\textwidth}
\includegraphics[width=\textwidth]{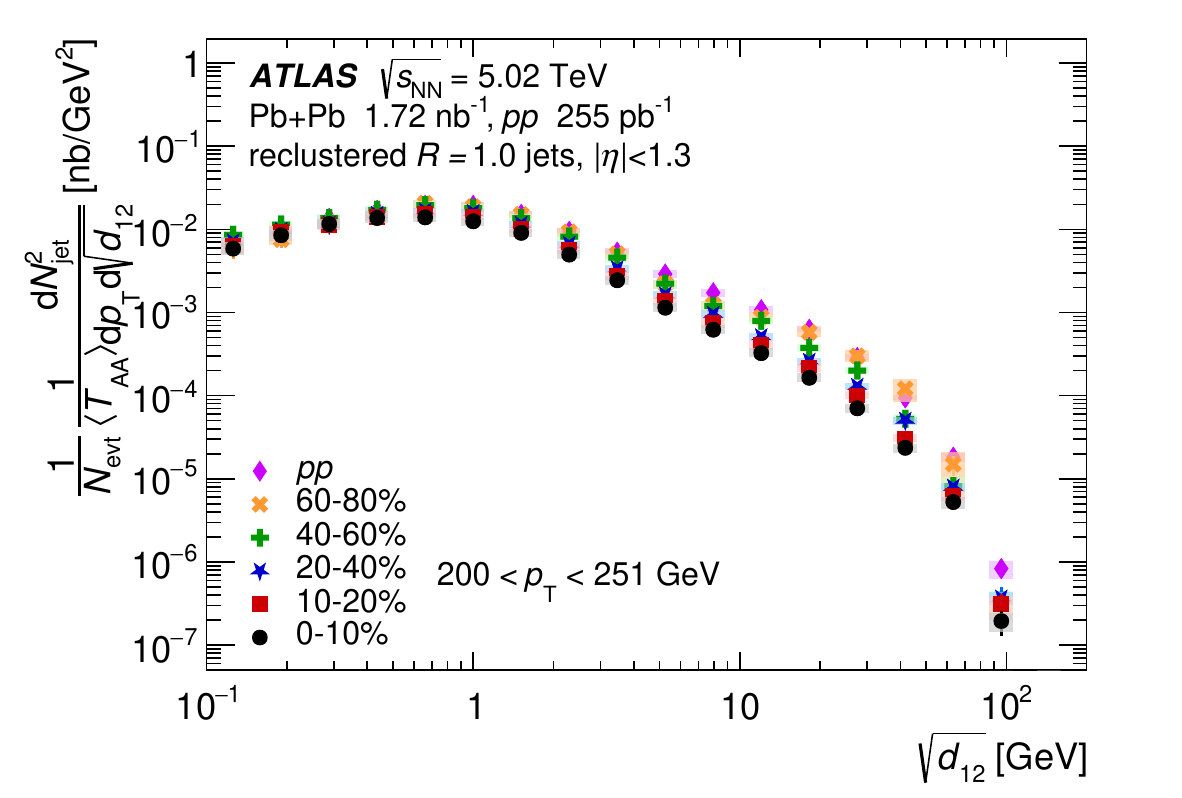}
\caption{}
\label{fig:yields_a}
\end{subfigure}
\begin{subfigure}[b]{0.49\textwidth}
\includegraphics[width=\textwidth]{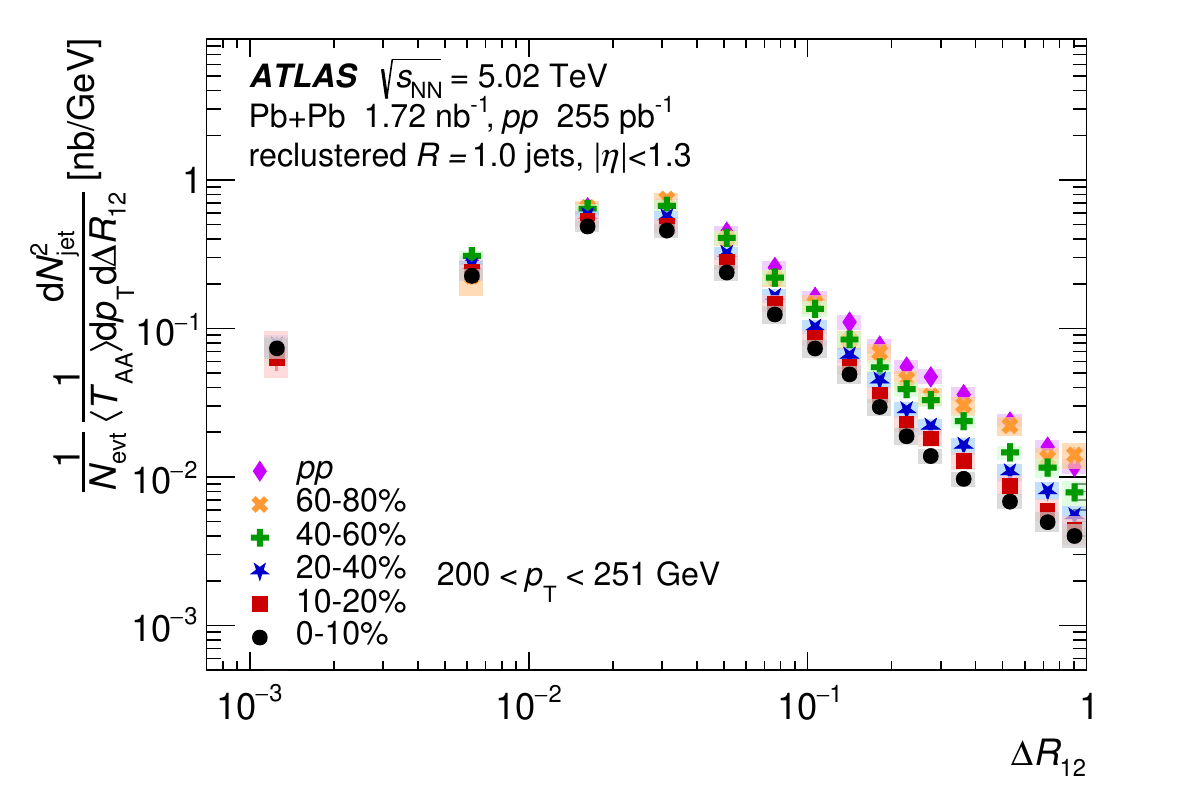}
\caption{}
\label{fig:yields_b}
\end{subfigure}
\caption{
Measured cross-sections in \pp\ collisions and $\langle\TAA\rangle$-normalized per-event yields in \PbPb collisions, evaluated as a function of (a)~\Donetwo and (b)~\dRonetwo, for reclustered $R=1.0$ jets in the \pt\ interval $200{-}251$~GeV.
Normalized yields are shown for five centrality bins of \PbPb collisions.
The vertical bars on the data points indicate statistical uncertainties, while the shaded boxes indicate systematic uncertainties. The shown systematic uncertainties do not include the uncertainties in the luminosity and $\langle\TAA\rangle$ determinations, which are correlated for all the data points and are listed in Section~\ref{sec:uncert}.
}
\end{figure}

The distributions of \Donetwo for \lR jets in the \pt interval $200{-}251$~\GeV measured in \pp\ collisions and five centrality bins of \PbPb\ collisions are shown in Figure~\ref{fig:yields_a}. All \Donetwo distributions have similar shape, rising at low \Donetwo\ values, reaching a maximum at $\Donetwo \approx 1$~GeV, and then decreasing. This shape was seen in the previous measurement with \lR jets~\cite{HION-2019-09}, albeit with the maximum shifted towards larger \Donetwo values as a consequence of using $R=0.2$ jets above a higher \pt threshold as the input to the reclustering.
The distributions of \Ronetwo for the same kinematic range and event selections as used for the \Donetwo distributions are shown in Figure~\ref{fig:yields_b}. All the \Ronetwo distributions feature the same trend:\ rising, reaching a maximum, and then decreasing with increasing \Ronetwo. The maximum is reached at $\Ronetwo \approx 0.02$, which is consistent with the value measured for
the angular separation of the two leading sub-jets, $r_\mathrm{g}$, in groomed
$R=0.4$ jets in the previous measurement~\cite{HION-2021-09} which used charged-particle tracks matched to calorimeter energy clusters to investigate the jet substructure.

\begin{figure}[t]
\begin{subfigure}[b]{0.49\textwidth}
\includegraphics[width=\textwidth]{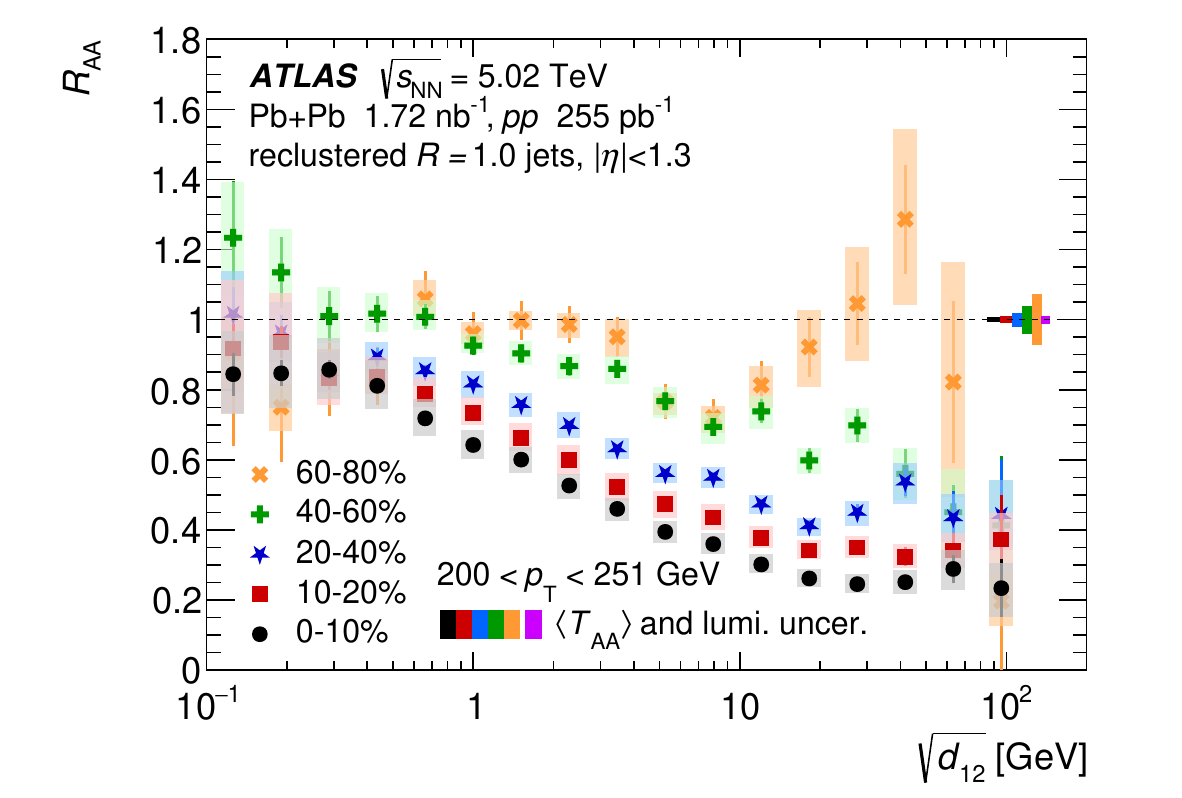}
\caption{}\label{fig:RAA_a}
\end{subfigure}
\begin{subfigure}[b]{0.49\textwidth}
\includegraphics[width=\textwidth]{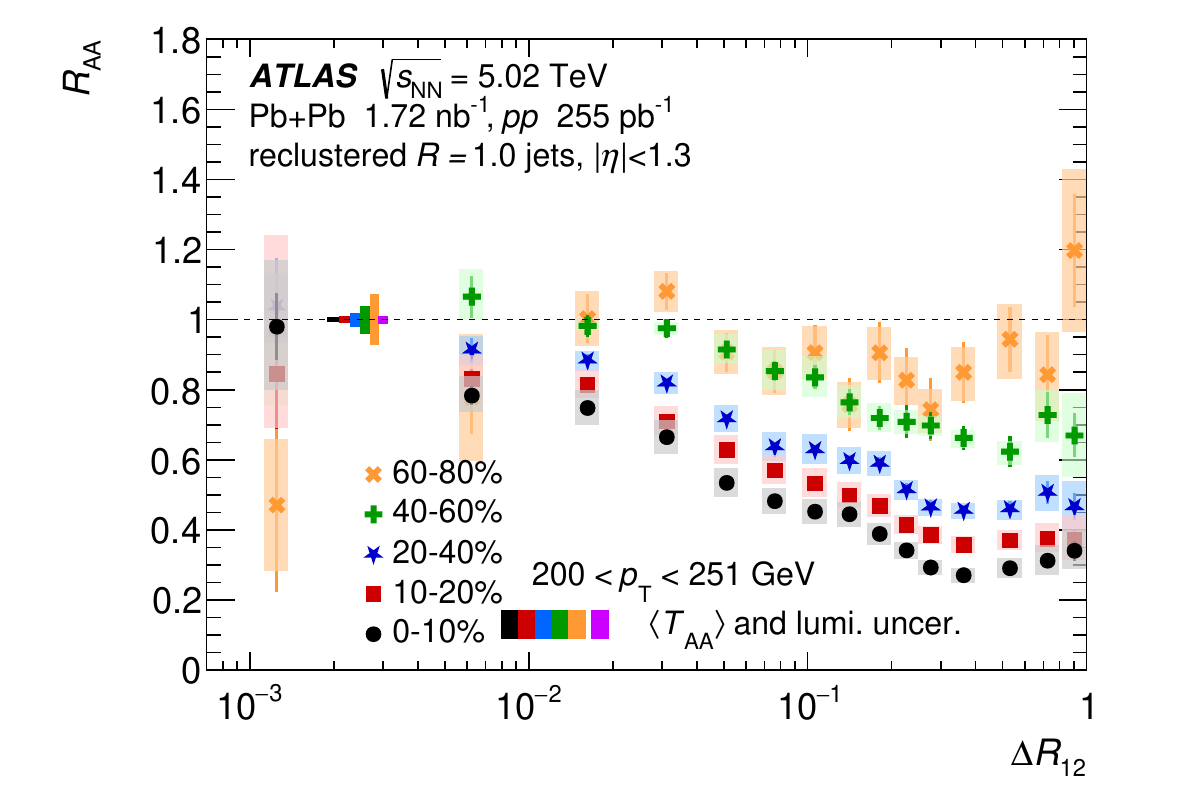}
\caption{}\label{fig:RAA_b}
\end{subfigure}
\caption{
The \RAA\ of $R=1.0$ jets evaluated as a function of (a)~\Donetwo and (b)~\dRonetwo in five centrality bins and the \pt\ interval $200{-}251$~GeV.
The vertical bars on the data points indicate statistical uncertainties, while the shaded boxes indicate systematic uncertainties. The coloured boxes at $\RAA\ = 1$ represent the fractional per-centrality uncertainty in $\langle\TAA\rangle$ and $pp$ luminosity (magenta), which both affect the overall normalization.
}
\end{figure}

To quantitatively assess the differences in the suppression of jets characterized by different values of substructure observables, the \RAA of \lR jets is evaluated as a function of \Donetwo and \Ronetwo in bins of the jet \pt and in five centrality intervals.
The \RAA evaluated as a function of \Donetwo in the jet \pt interval $200{-}251$~\GeV is shown in Figure~\ref{fig:RAA_a}.
For values of $\Donetwo \gtrsim 0.7$~\GeV, one can clearly see that the suppression increases from peripheral to central collisions.
The \RAA gradually decreases with increasing \Donetwo in the range $0.7 \lesssim \Donetwo \lesssim 20$~\GeV in all centrality bins except the one for $60{-}80\%$ peripheral collisions.
In the $0{-}10\%$ central collisions, the \RAA value decreases from approximately 0.8 to 0.2, implying significantly stronger suppression of \lR jets with a larger $\kt$ splitting scale.
In the $60{-}80\%$ peripheral collisions, the distribution's downward trend with increasing \Donetwo is not pronounced, and the \RAA fluctuates around a value of approximately 0.8.
For $\Donetwo \lesssim 0.7$~\GeV, the \RAA is greater than 0.8 and it is not possible to discern a significant difference in \RAA value between different centrality bins.

The \RAA evaluated as a function of \Ronetwo for the same kinematic range and centrality bins as in the case of \Donetwo is shown in Figure~\ref{fig:RAA_b}. In the range $0.01 \lesssim \Ronetwo \lesssim 0.2$, the \RAA decreases in all centrality bins except for $60{-}80\%$. For $\Ronetwo \lesssim 0.01$, it is not possible to discern a significant difference among centralities because of large uncertainties. For $\Ronetwo \gtrsim 0.3$, the \RAA values are consistent with being constant within the systematic uncertainties, implying no dependence of the jet suppression on the angular separation of the two sub-jets. This flattening is seen in all centrality selections except for the most peripheral collisions, where the overall jet suppression is small and \RAA is more susceptible to statistical fluctuations.
The onset of increasing \RAA with decreasing values of \DRonetwo may be partially connected with the suppression difference between narrow and wide jets, but should also be connected with the onset of coherent radiation and should therefore allow one to directly constrain the values of $\lres$ and $\theta_\text{c}$ as discussed, e.g., in Ref.~\cite{Kudinoor:2025ilx}. Similar trends in \RAA\ as a function of \Donetwo\ and \dRonetwo\ are also observed for higher-\pt\ \lR\ jets, although with larger statistical uncertainties.

\begin{figure}[t]
\begin{subfigure}[b]{0.49\textwidth}
\includegraphics[width=\textwidth]{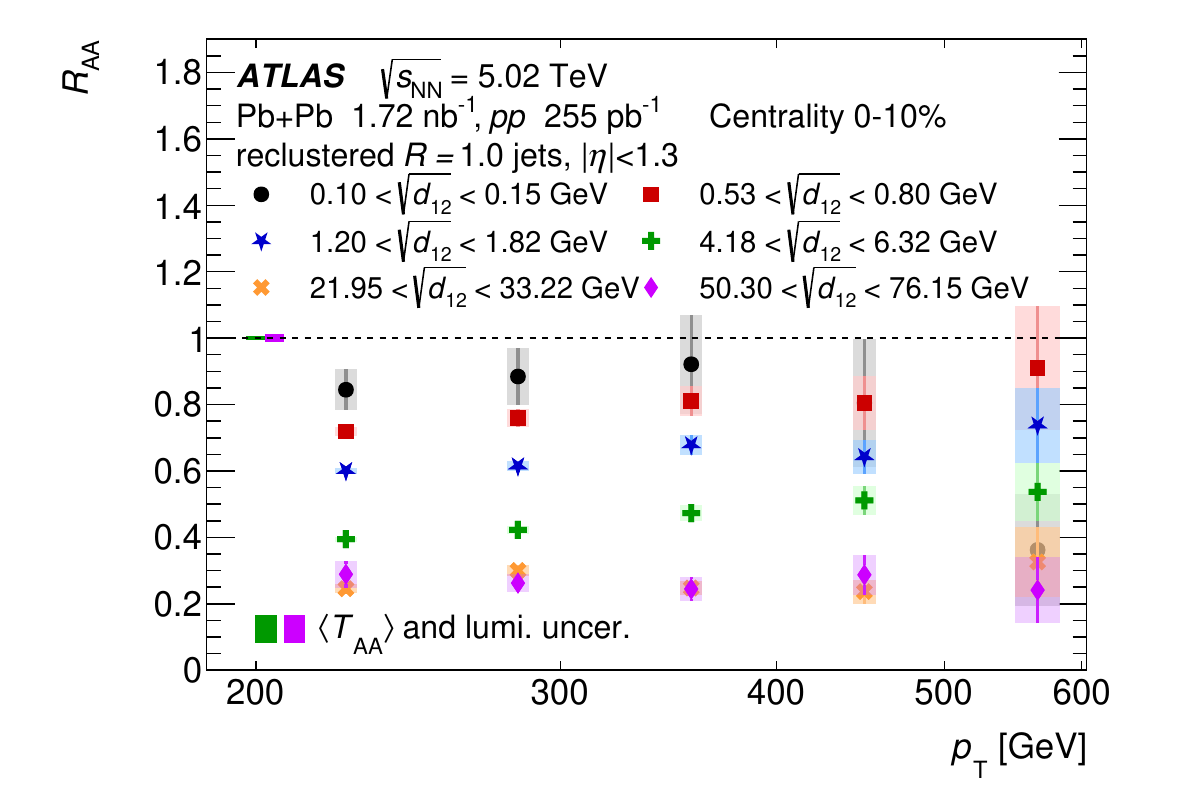}
\caption{}\label{fig:RAAvspT_a}
\end{subfigure}
\begin{subfigure}[b]{0.49\textwidth}
\includegraphics[width=\textwidth]{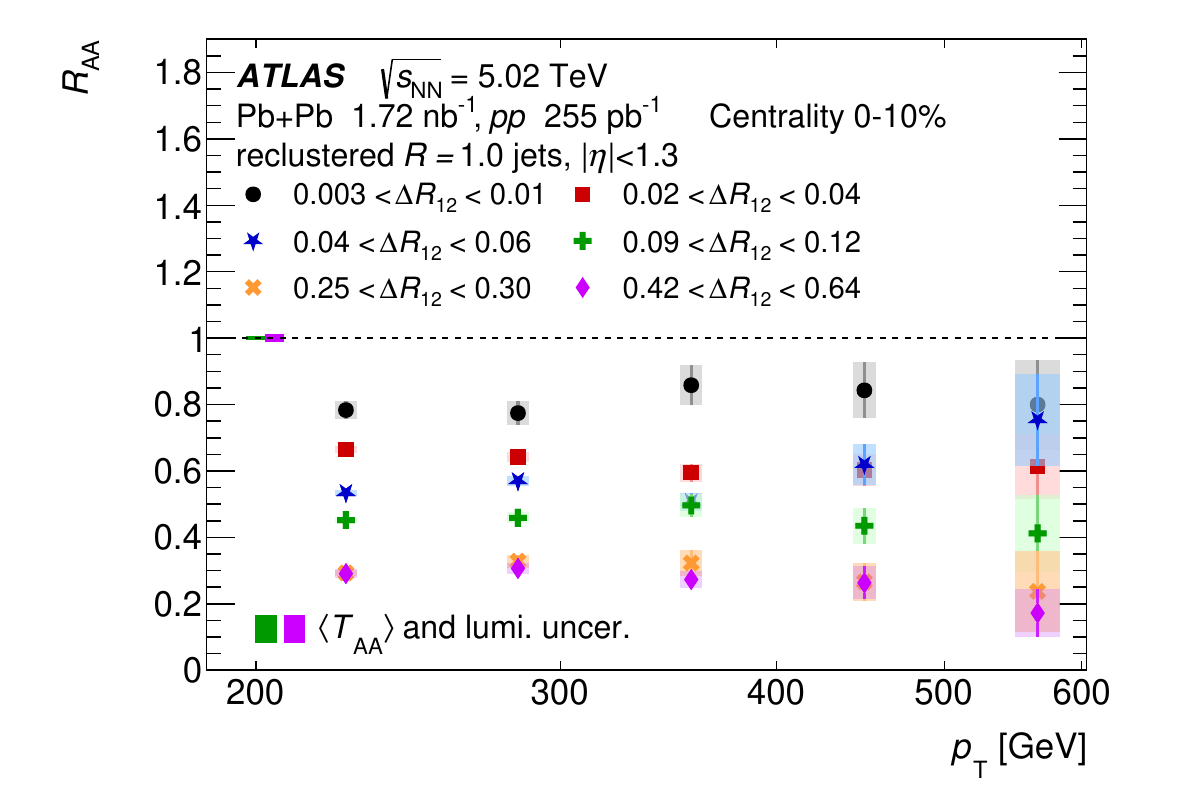}
\caption{}\label{fig:RAAvspT_b}
\end{subfigure}
\caption{
The \RAA\ of $R=1.0$ jets in the most central 0--10\% collisions evaluated as a function of jet \pt in six intervals of (a)~\Donetwo and (b)~\dRonetwo.
The vertical bars on the data points indicate statistical uncertainties, while the shaded boxes indicate systematic uncertainties. The coloured boxes at $\RAA\ = 1$ represent the fractional per-centrality uncertainty in $\langle\TAA\rangle$ and $pp$ luminosity (magenta), which both affect the overall normalization.
}
\end{figure}

To quantify the jet \pt dependence, the \RAA is evaluated in $0{-}10\%$ central collisions as a function of jet \pt for several \Donetwo and \Ronetwo bins as shown in Figure~\ref{fig:RAAvspT_a} and Figure~\ref{fig:RAAvspT_b}, respectively. The substructure-observable bins are selected to quantify the \RAA in three distinct regions:\ the region with low \Donetwo and \Ronetwo values ($\Donetwo \lesssim 0.5$~GeV, $\Ronetwo \lesssim 0.01$); the region with intermediate \Donetwo and \Ronetwo values ($0.5 \lesssim \Donetwo \lesssim 20$~GeV, $ 0.01 \lesssim \Ronetwo \lesssim 0.3$) where \RAA decreases with increasing \Donetwo and \Ronetwo; and the region with high \Donetwo and \Ronetwo values ($\Donetwo \gtrsim 20$~GeV, $ \Ronetwo \gtrsim 0.3$) where \RAA reaches a constant value. Figures~\ref{fig:RAAvspT_a} and ~\ref{fig:RAAvspT_b} show that the \RAA is independent of jet \pt in the region with low values of the substructure observables. The \RAA is also independent of jet \pt in the region with high values of the substructure observables. In the region with intermediate values of the substructure observables, the \RAA tends to increase with increasing jet \pt when evaluated in bins of \Donetwo. This tendency was confirmed quantitatively by fitting the distribution with a linear function. When evaluated in bins of \Ronetwo, any such increase is very mild or absent. This quantification of the jet \pt dependence of \RAA may further help to constrain the mechanisms responsible for the significant dependence of jet suppression on jet substructure.

\begin{figure}[t]
\centerline{
\includegraphics[width=0.79\textwidth]{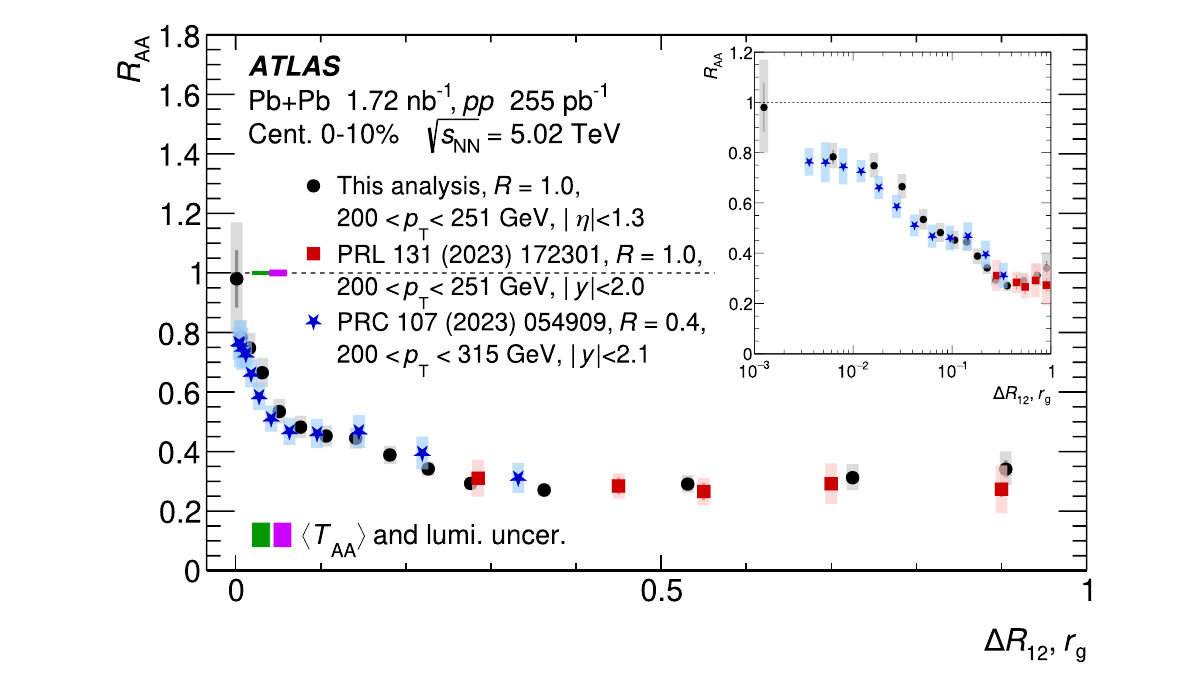}
}
\caption{The \RAA\ of $R=1.0$ jets evaluated as a function of $\Ronetwo$ compared with measurements from Ref.~\cite{HION-2019-09} ($R=1.0$ jets with $R=0.2$ constituents used to define sub-jets) and Ref.~\cite{HION-2021-09} ($R=0.4$ jets with charged-particle tracks matched to calorimeter energy clusters used to define sub-jets).
Results are shown in the \pt\ interval $200{-}251$~GeV for this measurement and the measurement in Ref.~\cite{HION-2019-09}, and $200{-}315$~GeV for the measurement in Ref.~\cite{HION-2021-09}. The inset displays the same distributions using a logarithmic scale, highlighting features at lower $\Ronetwo$ values. The vertical bars on the data points indicate statistical uncertainties, while the shaded boxes indicate systematic uncertainties. The coloured boxes at $\RAA\ = 1$ represent the fractional uncertainty in $\langle\TAA\rangle$ in 0--10\% \PbPb\ collisions (green) and $pp$ luminosity (magenta), which both affect the overall normalization.
}
\label{fig:RAAComparison}
\end{figure}

The results presented in this Letter are also compared with results from previous studies published in Refs.~\cite{HION-2019-09,HION-2021-09}. This comparison is shown in Figure~\ref{fig:RAAComparison}.
The \RAA\ evaluated as a function of \Ronetwo is compared with the \RAA\ of reclustered $R=1.0$ jets where the $R=0.2$ sub-jets reconstructed in the calorimeter are used to study the jet substructure~\cite{HION-2019-09}, and with the \RAA\ of $R=0.4$ jets where the substructure is quantified using tracks matched to calorimeter energy clusters~\cite{HION-2021-09}. The latter measurement evaluated a quantity called $r_\mathrm{g}$, which equates with \Ronetwo.
While the jet definitions and kinematic requirements differ between the measurements, it is still possible to directly compare the reported trends.
The values of \RAA measured in the flattening region in this Letter are consistent with those reported in Ref.~\cite{HION-2019-09}. The new measurement is also consistent with Ref.~\cite{HION-2021-09} in the region where \RAA decreases with increasing \Ronetwo ($r_\mathrm{g}$).

In the region where \RAA decreases, both this measurement and the measurement in Ref.~\cite{HION-2021-09} show the possible presence of an inflection point near $\Ronetwo \approx 0.1$.
Such an inflection point might be connected with the presence of competing mechanisms responsible for the change in \RAA. The significance of the presence of an inflection point was determined by fitting a third-order polynomial, and it was found to be 2.8 and 3.6 standard deviations for this measurement and the measurement in Ref.~\cite{HION-2021-09}, respectively.
Evidence for the presence of this inflection point may bring important information for modelling substructure-dependent jet suppression.
As a whole, the comparisons presented in Figure~\ref{fig:RAAComparison} show that the new measurement allows one to fully quantify
the onset of the increase in \RAA
with decreasing angular separation of the two sub-jets.

\begin{figure}[t]
\centerline{
\includegraphics[width=0.79\textwidth]{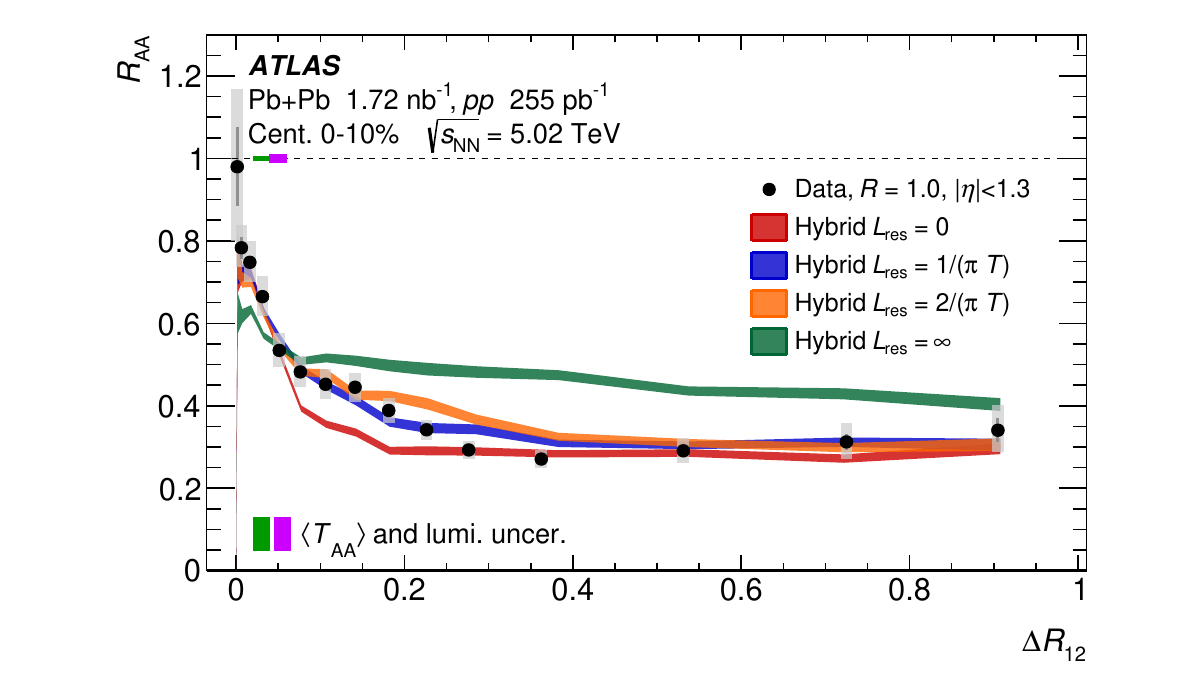}
}

\caption{
The \RAA\ of $R=1.0$ jets as a function of $\Ronetwo$ is compared to theoretical calculations from the Hybrid model~\cite{Kudinoor:2025ilx,Casalderrey-Solana:2014bpa,Hulcher:2017cpt}. Results are shown for $200{-}251$~GeV jets with $|\eta|<1.3$ in 0--10\% \PbPb\ collisions. The theory is displayed as coloured bands corresponding to resolution lengths, $L_{\mathrm{res}}=0$, $1/(\pi T)$, $2/(\pi T)$, and $L_{\mathrm{res}}\!\to\!\infty$. The vertical bars on the data points indicate statistical uncertainties, while the shaded boxes indicate systematic uncertainties. The coloured boxes at $\RAA=1$ represent the fractional uncertainties in $\langle\TAA\rangle$ (green) and the $pp$ luminosity (magenta), which affect the overall normalisation.
}
\label{fig:hybrid_comparison}
\end{figure}

A comparison of the data to theoretical calculations from the Hybrid model ~\cite{Kudinoor:2025ilx,Casalderrey-Solana:2014bpa,Hulcher:2017cpt} is presented in Fig.~\ref{fig:hybrid_comparison}. In the Hybrid model framework, $L_{\mathrm{res}}$ represents the earlier discussed resolution length and parametrizes the ability of the medium to resolve the substructure of the jet. The measurement disfavors extreme resolution lengths, namely $L_{\mathrm{res}}=0$ and $L_{\mathrm{res}} \rightarrow \infty$, where the exclusion of the latter case was already discussed in ~\cite{Kudinoor:2025ilx}. The data exhibit a mild preference for $L_{\mathrm{res}} = 1/(\pi T)$ over $L_{\mathrm{res}} = 2/(\pi T)$. This comparison demonstrates that the presented measurement has a substantial constraining power on the medium resolution scale and thus should help to quantify the impact of color coherence on measured jet suppression.

\FloatBarrier


%

%
\section{Conclusion}

This Letter presents a measurement of the jet substructure-dependent
suppression of \lR jets performed with the ATLAS detector at the LHC using 1.72~nb$^{-1}$ and 255~pb$^{-1}$ samples of \PbPb\ and \pp\ collision data, respectively, collected at a centre-of-mass energy of 5.02~TeV per colliding nucleon pair. The jets were reconstructed using the \antikt algorithm with $R=1.0$ by reclustering $R=0.2$ jets with $\pt>35$~\GeV following the strategy used in a recently published ATLAS analysis. To study the substructure of \lR jets, charged-particle tracks were used, allowing a definition of the two leading sub-jets, their angular separation ($\Ronetwo$), and \kt\ splitting scale ($\Donetwo$). The measurement is corrected to the particle level and presented in bins of jet \pt and collision centrality for jets with $|\eta|<1.3$ and $\pt$ above 200~\GeV.

The jet \RAA\ in the 0--10~\% central collisions evaluated as a function of $\Donetwo$ gradually decreases with increasing $\Donetwo$, from a value of approximately 0.8 at $\Donetwo \approx 0.7$~\GeV to 0.2 at $\Donetwo \approx 20$~\GeV, implying significantly stronger suppression of \lR jets with larger \kt splitting scale. The jet \RAA evaluated as a function of $\Ronetwo$ gradually decreases for $\Ronetwo$ in the range $0.01{-}0.2$ and then remains consistent with a constant value for $\Ronetwo \gtrsim 0.2$. A precise evaluation of the onset of $\Ronetwo$- and $\Donetwo$-dependent jet suppression will bring new insights into the role of jet substructure in jet quenching and, in particular, will help to quantitatively constrain the impact of colour coherence on measured jet suppression.


%

%
\section*{Acknowledgements}
%

%
%

%
%

We thank CERN for the very successful operation of the LHC and its injectors, as well as the support staff at
CERN and at our institutions worldwide without whom ATLAS could not be operated efficiently.

The crucial computing support from all WLCG partners is acknowledged gratefully, in particular from CERN, the ATLAS Tier-1 facilities at TRIUMF/SFU (Canada), NDGF (Denmark, Norway, Sweden), CC-IN2P3 (France), KIT/GridKA (Germany), INFN-CNAF (Italy), NL-T1 (Netherlands), PIC (Spain), RAL (UK) and BNL (USA), the Tier-2 facilities worldwide and large non-WLCG resource providers. Major contributors of computing resources are listed in Ref.~\cite{ATL-SOFT-PUB-2025-001}.

We gratefully acknowledge the support of ANPCyT, Argentina; YerPhI, Armenia; ARC, Australia; BMWFW and FWF, Austria; ANAS, Azerbaijan; CNPq and FAPESP, Brazil; NSERC, NRC and CFI, Canada; CERN; ANID, Chile; CAS, MOST and NSFC, China; Minciencias, Colombia; MEYS CR, Czech Republic; DNRF and DNSRC, Denmark; IN2P3-CNRS and CEA-DRF/IRFU, France; SRNSFG, Georgia; BMFTR, HGF and MPG, Germany; GSRI, Greece; RGC and Hong Kong SAR, China; ICHEP and Academy of Sciences and Humanities, Israel; INFN, Italy; MEXT and JSPS, Japan; CNRST, Morocco; NWO, Netherlands; RCN, Norway; MNiSW, Poland; FCT, Portugal; MNE/IFA, Romania; MSTDI, Serbia; MSSR, Slovakia; ARIS and MVZI, Slovenia; DSI/NRF, South Africa; MICIU/AEI, Spain; SRC and Wallenberg Foundation, Sweden; SERI, SNSF and Cantons of Bern and Geneva, Switzerland; NSTC, Taipei; TENMAK, T\"urkiye; STFC/UKRI, United Kingdom; DOE and NSF, United States of America.

Individual groups and members have received support from BCKDF, CANARIE, CRC and DRAC, Canada; CERN-CZ, FORTE and PRIMUS, Czech Republic; COST, ERC, ERDF, Horizon 2020, ICSC-NextGenerationEU and Marie Sk{\l}odowska-Curie Actions, European Union; Investissements d'Avenir Labex, Investissements d'Avenir Idex and ANR, France; DFG and AvH Foundation, Germany; Herakleitos, Thales and Aristeia programmes co-financed by EU-ESF and the Greek NSRF, Greece; BSF-NSF and MINERVA, Israel; NCN and NAWA, Poland; La Caixa Banking Foundation, CERCA Programme Generalitat de Catalunya and PROMETEO and GenT Programmes Generalitat Valenciana, Spain; G\"{o}ran Gustafssons Stiftelse, Sweden; The Royal Society and Leverhulme Trust, United Kingdom.

In addition, individual members wish to acknowledge support from Armenia: Yerevan Physics Institute (FAPERJ); CERN: European Organization for Nuclear Research (CERN DOCT); Chile: Agencia Nacional de Investigaci\'on y Desarrollo (FONDECYT 1230812, FONDECYT 1240864, Fondecyt 3240661); China: Chinese Ministry of Science and Technology (MOST-2023YFA1605700, MOST-2023YFA1609300), National Natural Science Foundation of China (NSFC - 12175119, NSFC 12275265); Czech Republic: Czech Science Foundation (GACR - 24-11373S), Ministry of Education Youth and Sports (ERC-CZ-LL2327, FORTE CZ.02.01.01/00/22\_008/0004632), PRIMUS Research Programme (PRIMUS/21/SCI/017); EU: H2020 European Research Council (ERC - 101002463); European Union: European Research Council (BARD No. 101116429, ERC - 948254, ERC 101089007), European Regional Development Fund (SMASH COFUND 101081355, SLO ERDF), Horizon 2020 Framework Programme (MUCCA - CHIST-ERA-19-XAI-00), European Union, Future Artificial Intelligence Research (FAIR-NextGenerationEU PE00000013), Horizon 2020 (EuroHPC - EHPC-DEV-2024D11-051), Italian Center for High Performance Computing, Big Data and Quantum Computing (ICSC, NextGenerationEU); France: Agence Nationale de la Recherche (ANR-21-CE31-0022, ANR-22-EDIR-0002); Germany: Baden-Württemberg Stiftung (BW Stiftung-Postdoc Eliteprogramme), Deutsche Forschungsgemeinschaft (DFG - 469666862, DFG - CR 312/5-2); China: Research Grants Council (GRF); Italy: Istituto Nazionale di Fisica Nucleare (ICSC, NextGenerationEU), Ministero dell'Università e della Ricerca (NextGenEU 153D23001490006 M4C2.1.1, NextGenEU I53D23000820006 M4C2.1.1, NextGenEU I53D23001490006 M4C2.1.1, SOE2024\_0000023); Japan: Japan Society for the Promotion of Science (JSPS KAKENHI JP22H01227, JSPS KAKENHI JP22H04944, JSPS KAKENHI JP22KK0227, JSPS KAKENHI JP23KK0245, JSPS KAKENHI JP24K23939); Norway: Research Council of Norway (RCN-314472); Poland: Ministry of Science and Higher Education (IDUB AGH, POB8, D4 no 9722), Polish National Science Centre (NCN 2021/42/E/ST2/00350, NCN OPUS 2023/51/B/ST2/02507, NCN OPUS nr 2022/47/B/ST2/03059, NCN UMO-2019/34/E/ST2/00393, UMO-2022/47/O/ST2/00148, UMO-2023/49/B/ST2/04085, UMO-2023/51/B/ST2/00920, UMO-2024/53/N/ST2/00869); Portugal: Foundation for Science and Technology (FCT); Spain: Ministry of Science and Innovation (MCIN \& NextGenEU PCI2022-135018-2, MICIN \& FEDER PID2021-125273NB, RYC2019-028510-I, RYC2020-030254-I, RYC2021-031273-I, RYC2022-038164-I); Sweden: Carl Trygger Foundation (Carl Trygger Foundation CTS 22:2312), Swedish Research Council (Swedish Research Council 2023-04654, VR 2021-03651, VR 2022-03845, VR 2022-04683, VR 2023-03403, VR 2024-05451), Knut and Alice Wallenberg Foundation (KAW 2018.0458, KAW 2022.0358, KAW 2023.0366); Switzerland: Swiss National Science Foundation (SNSF - PCEFP2\_194658); United Kingdom: Leverhulme Trust (Leverhulme Trust RPG-2020-004), Royal Society (NIF-R1-231091); United States of America: U.S. Department of Energy (ECA DE-AC02-76SF00515), Neubauer Family Foundation.

%
%


%

%
%
%
\printbibliography
\clearpage
 
\begin{flushleft}
\hypersetup{urlcolor=black}
{\Large The ATLAS Collaboration}

\bigskip

\AtlasOrcid[0000-0002-6665-4934]{G.~Aad}$^\textrm{\scriptsize 104}$,
\AtlasOrcid[0000-0001-7616-1554]{E.~Aakvaag}$^\textrm{\scriptsize 17}$,
\AtlasOrcid[0000-0002-5888-2734]{B.~Abbott}$^\textrm{\scriptsize 123}$,
\AtlasOrcid[0000-0002-0287-5869]{S.~Abdelhameed}$^\textrm{\scriptsize 119a}$,
\AtlasOrcid[0000-0002-1002-1652]{K.~Abeling}$^\textrm{\scriptsize 55}$,
\AtlasOrcid[0000-0001-5763-2760]{N.J.~Abicht}$^\textrm{\scriptsize 49}$,
\AtlasOrcid[0000-0002-8496-9294]{S.H.~Abidi}$^\textrm{\scriptsize 30}$,
\AtlasOrcid[0009-0003-6578-220X]{M.~Aboelela}$^\textrm{\scriptsize 45}$,
\AtlasOrcid[0000-0002-9987-2292]{A.~Aboulhorma}$^\textrm{\scriptsize 36e}$,
\AtlasOrcid[0000-0001-5329-6640]{H.~Abramowicz}$^\textrm{\scriptsize 157}$
% ...
\bigskip
\\

% ...
$^{17}$Department for Physics and Technology, University of Bergen, Bergen; Norway.\\
% ...
$^{30}$Physics Department, Brookhaven National Laboratory, Upton NY; United States of America.\\
% ...
$^{36}$$^{(a)}$Facult\'e des Sciences Ain Chock, Universit\'e Hassan II de Casablanca;$^{(b)}$Facult\'{e} des Sciences, Universit\'{e} Ibn-Tofail, K\'{e}nitra;$^{(c)}$Facult\'e des Sciences Semlalia, Universit\'e Cadi Ayyad, LPHEA-Marrakech;$^{(d)}$LPMR, Facult\'e des Sciences, Universit\'e Mohamed Premier, Oujda;$^{(e)}$Facult\'e des sciences, Universit\'e Mohammed V, Rabat;$^{(f)}$Institute of Applied Physics, Mohammed VI Polytechnic University, Ben Guerir; Morocco.\\
% ...
$^{45}$Physics Department, Southern Methodist University, Dallas TX; United States of America.\\
% ...
$^{49}$Fakult\"{a}t Physik , Technische Universit{\"a}t Dortmund, Dortmund; Germany.\\
% ...
$^{55}$II. Physikalisches Institut, Georg-August-Universit\"{a}t G\"ottingen, G\"ottingen; Germany.\\
% ...
$^{104}$CPPM, Aix-Marseille Universit\'e, CNRS/IN2P3, Marseille; France.\\
% ...
$^{119}$$^{(a)}$New York University Abu Dhabi, Abu Dhabi;$^{(b)}$United Arab Emirates University, Al Ain; United Arab Emirates.\\
% ...
$^{123}$Homer L. Dodge Department of Physics and Astronomy, University of Oklahoma, Norman OK; United States of America.\\
% ...
$^{157}$Raymond and Beverly Sackler School of Physics and Astronomy, Tel Aviv University, Tel Aviv; Israel.\\
% ...
$^{a}$ Also at Affiliated with an institute formerly covered by a cooperation agreement with CERN.\\
$^{b}$ Also at An-Najah National University, Nablus; Palestine.\\
$^{c}$ Also at Borough of Manhattan Community College, City University of New York, New York NY; United States of America.\\
$^{d}$ Also at Center for Interdisciplinary Research and Innovation (CIRI-AUTH), Thessaloniki; Greece.\\
$^{e}$ Also at Centre of Physics of the Universities of Minho and Porto (CF-UM-UP); Portugal.\\
$^{f}$ Also at CERN, Geneva; Switzerland.\\
% ...

\end{flushleft}


%

%
%
%
\clearpage

\end{document}